\documentclass[pra,twocolumn,showpacs]{revtex4}
\usepackage{epsfig}
\usepackage{graphicx}
\usepackage{amsmath}
\usepackage{natbib}
\newcommand{\bn}{\begin{eqnarray}}
\newcommand{\en}{\end{eqnarray}}
\newcommand{\eml}{\end{multline}}
\newcommand{\bml}{\begin{multline}}

\begin{document}

\title {Twisting tensor and spin squeezing}
 \author{Tom\'{a}\v{s} Opatrn\'{y}}
 \affiliation{Optics Department, Faculty of Science, Palack\'{y} University, 17. Listopadu 12,
 77146 Olomouc, Czech Republic}

\date{\today }
\begin{abstract}
A unified tensor description of quadratic spin squeezing interactions is proposed, covering the single- and two-axis twisting as  special cases of a general scheme. A closed set of equations of motion of the first moments and variances is derived in Gaussian approximation and their solutions are discussed from the prospect of fastest squeezing generation. It turns out that the optimum rate of squeezing generation is governed by the difference between the largest and the smallest eigenvalues of the twisting tensor. A cascaded optical interferometer with Kerr nonlinear media is proposed as one of possible realizations of the general scheme.   
\end{abstract}
\pacs{42.50.Lc, 37.25.+k, 03.75.Dg, 03.75.Gg }
%
%

\maketitle

\section{Introduction}
Suppressed noise in two-mode multi-particle systems known as ``spin squeezing'' introduced by Kitagawa and Ueda \cite{Kitagawa} is an essential tool in quantum metrology protocols \cite{Wineland1994,Lloyd}. The interferometric schemes utilizing this effect cover broad area of possible physical systems, ranging from collective spins of neutral atoms interacting by collisions \cite{Micheli2003,Esteve2008,Gross2010,Riedel}, atoms interacting with light by Faraday rotation and ac-Stark shift \cite{AtomLight}, atoms interacting by Rydberg blockade \cite{Rydberg}, polarized light \cite{Korolkova},  to Bose-Einstein condensates (BEC) in double-well potentials (bosonic Josephson junctions) \cite{Vardi2001,Pezze,Diaz2012a}. Typically, the preparation of spin squeezed states is based on nonlinear inter-particle interactions. 
In terms of the collective ``spin'' operators $J$, the procedures have been classified as ``one-axis twisting'' (OAT) with a term $J_z^2$, and ``two-axis counter-twisting'' (TACT) with a term $J_x^2-J_y^2$, the TACT being shown to be more efficient to produce highly squeezed states \cite{Kitagawa}. Recently, a scheme has been proposed to combine a sequence of OAT and spin rotations to an effectively TACT procedure \cite{Liu2011}. Efficient preparation of spin-squeezed states has become an objective of various optimized procedures \cite{Diaz2012}. Here I show that any quadratic interaction in the collective spin can be described by means of a twisting tensor, encompassing the OAT and TACT as special cases. Equations of motion for the first and second moments in the Gaussian approximation are used to show how squeezing is generated in various cases of the twisting tensor. At the initial stage, the maximum squeezing rate only depends on the difference between the maximum and minimum eigenvalues of the twisting tensor.
For certain times, deviations from the optimum squeezing rate can be compensated by suitable rotations. The results are applicable for optimizing strategies of interferometric measurements with various nonlinear media. 

The paper is organized as follows. In Sec. \ref{secHam} the system Hamiltonian and equations of motion are derived, in Sec. \ref{secPhys} possible schemes for physical realization are mentioned, in Sec. \ref{secRate} the rate of squeezing generation is studied, in Sec. \ref{secSolution} approximate solutions of the equations of motion are given, in Sec. \ref{secOptimum} the conditions for generating squeezing at maximum rate are found, and a conclusion is given in  Sec. \ref{secConcl}.

\section{System Hamiltonian and equations of motion}
\label{secHam}
Consider a two-mode bosonic system described by annihilation operators $a$ and $b$ with total number of particles $N$ conserved. The dynamics can be expressed by operator $\vec{J}$ defined as 
\begin{eqnarray}
J_x &=& \frac{1}{2}(a^{\dag}b+ab^{\dag}), \\
J_y &=& \frac{1}{2i}(a^{\dag}b-ab^{\dag}), \\ 
J_z &=& \frac{1}{2}(a^{\dag}a-b^{\dag}b), 
\end{eqnarray}
with
$N = a^{\dag}a+b^{\dag}b$. The components of $\vec{J}$ satisfy the angular momentum  commutation relations $[J_x,J_y]=iJ_z$,  $[J_y,J_z]=iJ_x$, and $[J_z,J_x]=iJ_y$.
Let the Hamiltonian be composed of $a$, $b$, $a^{\dag}$, and  $b^{\dag}$ such that in each term the same number of creation and annihilation operators occurs (total number of particles is conserved), and the highest power of each operator is 2. The Hamiltonian then can be written as
\begin{eqnarray}
H = \omega_k J_k +\chi_{kl} J_k J_l + f(N), 
\label{Ham}
\end{eqnarray}
where 
$\omega_k$ and $J_k$ transform as vectors and $\chi_{kl}=\chi_{lk}$ transforms as a tensor under O(3) rotations. Here $k,l \in (x,y,z)$, and the Einstein summation is used.  In Eq. (\ref{Ham}), $f(N)$ is a linear or quadratic function of the total particle number,  generating an unimportant overall phase.
Let us call $\chi$ the {\em twisting tensor} and note that the special case of $\chi_{k,l}=0$ for $k\neq z$ or $l \neq z$, $\chi_{zz}\neq 0$ corresponds to the OAT scenario, and the case $\chi_{xx}=-\chi_{yy}\neq 0$, $\chi_{kl}=0$ otherwise, corresponds to the TACT scenario of \cite{Kitagawa}. Since $J_x^2+J_y^2+J_z^2=\frac{N}{2}(\frac{N}{2}+1)$, addition of an arbitrary multiple of unit matrix to $\chi$ can be absorbed in the unimportant term $f(N)$. Therefore, any diagonal $\chi$ in which one element is exactly in the middle of the remaining two elements also corresponds to the TACT. Hamiltonian (\ref{Ham}) specified by three parameters $\Omega, V, W$ as $\omega_z=\Omega$, $\chi_{xx}= V+W$, $\chi_{yy} = V-W$, with $\omega_k = \chi_{kl}=0$ otherwise,  corresponds to the Lipkin-Meshkov-Glick (LMG) \cite{Lipkin} model which was introduced as a solvable model of atomic nucleus and serves as a paradigm to study quantum phase transitions \cite{LMG-phase}.

\begin{figure}
\centerline{\epsfig{file=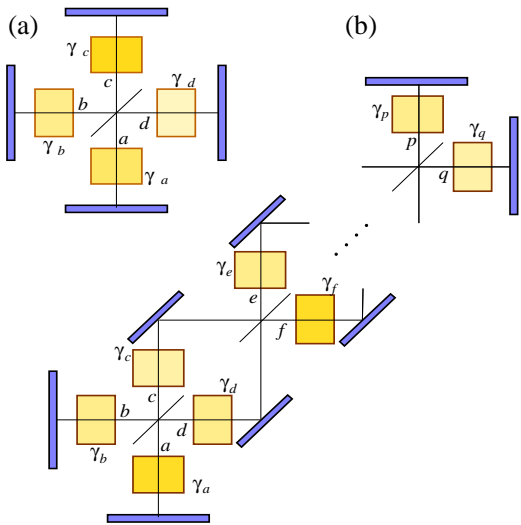,scale=0.40}}
\caption{\label{f-scheme1} (color online). A possible scheme realizing the interaction. (a) Modes of two crossed resonators mix at a balanced beam splitter according to $c=(a+b)/\sqrt{2}$, and $d=(a-b)/\sqrt{2}$, and each of the four beams propagates through a nonlinear medium. This setup leads to the Hamiltonian of Eq. (\ref{Ham1BS}). (b) Chaining of the mode mixing plus nonlinearities leads to the general form of Hamiltonian (\ref{Ham}).}
\end{figure}

Using the Heisenberg equations of motion, $i\dot{A} = [A,H]$, and calculating the mean values of the operators, we arrive at the equations for ${\cal J}_j \equiv \langle J_j\rangle$ 
\begin{eqnarray}
\dot {\cal J}_j = \epsilon_{jkl}[\omega_k {\cal J}_l+ 2\chi_{kn}({\cal J}_n {\cal J}_l+ V_{nl})] ,
\label{dotJ}
\end{eqnarray}
where $V$ is the variance tensor, 
\begin{eqnarray}
V_{nl} \equiv \frac{1}{2}\left\langle (J_n-{\cal J}_n)(J_l-{\cal J}_l) +(J_l-{\cal J}_l) (J_n-{\cal J}_n)\right\rangle .
\end{eqnarray}
The equations for $\dot{V}_{nl}$ can be obtained in a similar way, however, in this case mean values of cubic terms $\langle J_k J_s J_p\rangle$ occur. Our approximation is based on the assumption that the distribution of the components $J_k$ is close to Gaussian for which all higher moments are functions of the first and second moments. In particular, we express the third moments as 
\begin{eqnarray}
\langle J_k J_s J_p\rangle \approx {\cal J}_k {\cal J}_s{\cal J}_p
+ {\cal J}_k V_{sp} + {\cal J}_s V_{pk} + {\cal J}_p V_{ks}.
\end{eqnarray}
Thus we find
\begin{eqnarray}
\dot{V}_{kl} &\approx& \left(\omega_j + 2\chi_{js}{\cal J}_s\right)
\left(\epsilon_{plj} V_{pk} + \epsilon_{pkj} V_{pl} \right)
\nonumber \\
& & + 2 \chi_{js} {\cal J}_p \left(\epsilon_{plj} V_{sk} + \epsilon_{pkj} V_{sl} \right) ,
\label{dotV}
\end{eqnarray}
where $\epsilon$ is the Levi-Civita symbol.
Equations (\ref{dotJ}) and (\ref{dotV}) form a closed set of 9 equations for 9 dynamical variables describing rotational and squeezing properties of the system. Note that a special case of this set for OAT with $\omega=(-\omega_x,0,0)$ and $\chi_{kl}=0$ for $k\neq z \neq l$ has been studied in \cite{Vardi2001} where the influence of the variances $V_{nl}$ on the first moments ${\cal J}_j$ in Eq. (\ref{dotJ}) has been interpreted as the ``Bogoliubov backreaction''.

\section{Physical realization}
\label{secPhys}
A simple scheme realizing nontrivial $\chi$ is in Fig. \ref{f-scheme1}(a). Two optical resonators are crossed and their fields are mixed by a balanced beam splitter. In each of the four branches a Kerr medium induces a phase shift proportional to the intensity of the field. Thus, e.g., light passing through the medium in branch $a$ picks up the phase $\gamma_a |a|^2$.
If the round-trip duration is $\Delta t$, the Hamiltonian can be written as
\begin{eqnarray}
 H= \frac{1}{\Delta t}\left\{ \left(\gamma_a+ \gamma_b + \gamma_c+\gamma_d\right)\frac{N}{2}\left(\frac{N}{2}+1 \right) \right. \nonumber \\
+ (\gamma_a+ \gamma_b) J_z^2 + (\gamma_c+ \gamma_d) J_x^2 \nonumber \\
\left. +(N-1)(\gamma_a- \gamma_b) J_z + (N-1)(\gamma_c- \gamma_d) J_x \right\}. 
\label{Ham1BS}
\end{eqnarray}
A special case of $\gamma_a = \gamma_b \equiv \gamma_z$, and $\gamma_c = \gamma_d \equiv \gamma_x$ reduces the Hamiltonian to 
\begin{eqnarray}
H= \frac{2}{\Delta t}\left( \gamma_z J_z^2 +\gamma_x  J_x^2 \right) +f(N).
\end{eqnarray} 
Note that already this simplest form of interaction  covers all the main categories of spin squeezing: OAT ($\gamma_x= 0$ or $\gamma_z= 0$), TACT ($\gamma_x= 2\gamma_z$ or $\gamma_z= 2\gamma_x$), or more general twistings (other relations between $\gamma_x$ and $\gamma_z$), as well as the LMG model [$\gamma_c=\gamma_d=W$, $\gamma_{a,b}=(W-V)/2 \pm \Omega/(2(N-1))$]. 
Although in the last expression the rotation frequency depends on $N$,
rotating terms of various $\omega_{k}$ independent of $N$ can be introduced by shifting the positions of the mirrors and/or by tuning the parameters of the beam splitter.
Any more general form of the twisting tensor, including off-diagonal terms, can be achieved by chaining the beam splitters and nonlinear zones as in Fig. \ref{f-scheme1}(b).  The model is general: rather than optical resonators one can assume, for example, two bosonic traps with several Josephson junctions and position-dependent nonlinearities induced by Feshbach resonances. Recently, a scheme with a ring BEC trap with spatially modulated nonlinearity has been proposed to realize TACT and other squeezing regimes as well as the LMG model \cite{OKD14}.

Depending on the particular physical realization, a specific decoherence or loss mechanism will limit performance of the scheme. For instance, in the optical scheme the absorption may become dominant, whereas in the scheme discussed in \cite{OKD14} we anticipate the inelastic atomic collisions to represent the most important limitation. 
Detailed discussion of the influence of particle losses and thermal noise for the OAT BEC schemes have been given in \cite{Losses}.
Expanding the model to cover general twisting interactions,
eqs. (\ref{dotJ}) and (\ref{dotV}) would then be generalized based on the corresponding master equation. These problems will be studied in a subsequent work.

\section{Squeezing rate}
\label{secRate}
Let us first choose the coordinate system such that the state is centered at the pole of the Bloch sphere with ${\cal J}_x = {\cal J}_y = 0$. Using (\ref{dotV}) we find $\dot{V}_{xx}$, $\dot{V}_{xy}$ and $\dot{V}_{yy}$ while expressing the variance matrix $V_{kl}$, $k,l=x,y$  as the rotated diagonal matrix of principal variances $V_{\pm}$, where
\begin{eqnarray}
V_{\pm} = \frac{V_{xx}+V_{yy}}{2}\pm \sqrt{V_{xy}^2 + \frac{\left( V_{xx}-V_{yy} \right)^2}{4}}.
\end{eqnarray}
Thus we find
\begin{eqnarray}
 \dot{V}_{\pm} = \pm 2{\cal J}_z \left[(\chi_{yy}-\chi_{xx}) \sin 2\alpha - 2 \chi_{xy} \cos 2\alpha \right]V_{\pm} ,
\end{eqnarray}
where $\alpha$ is the orientation angle of the squeezed state. The optimum rate occurs for $\alpha$ satisfying
\begin{eqnarray}
 \tan 2\alpha = \frac{\chi_{xx}-\chi_{yy}}{2 \chi_{xy}}
 \label{tan2a}
\end{eqnarray}
for which
\begin{eqnarray}
 \dot{V}_{\pm}^{\rm (opt)} = \pm Q V_{\pm} ,
\end{eqnarray}
where
\begin{eqnarray}
 Q \equiv 2|{\cal J}_z| \sqrt{(\chi_{xx}-\chi_{yy})^2 + 4\chi_{xy}^2} 
\label{rateQ}
\end{eqnarray}
is the optimum squeezing rate. 

\begin{figure}
\centerline{\epsfig{file=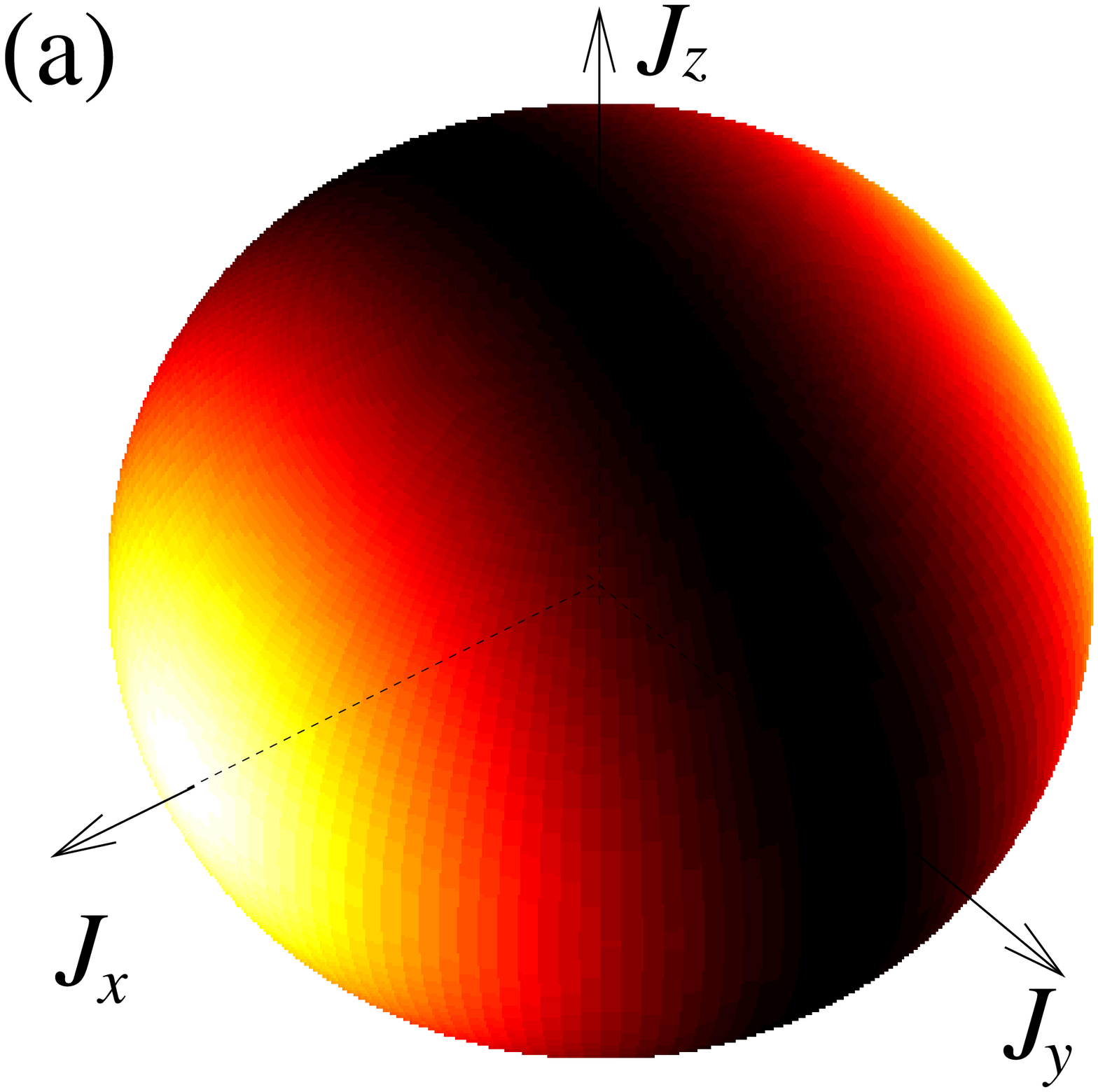,scale=0.18} \quad \epsfig{file=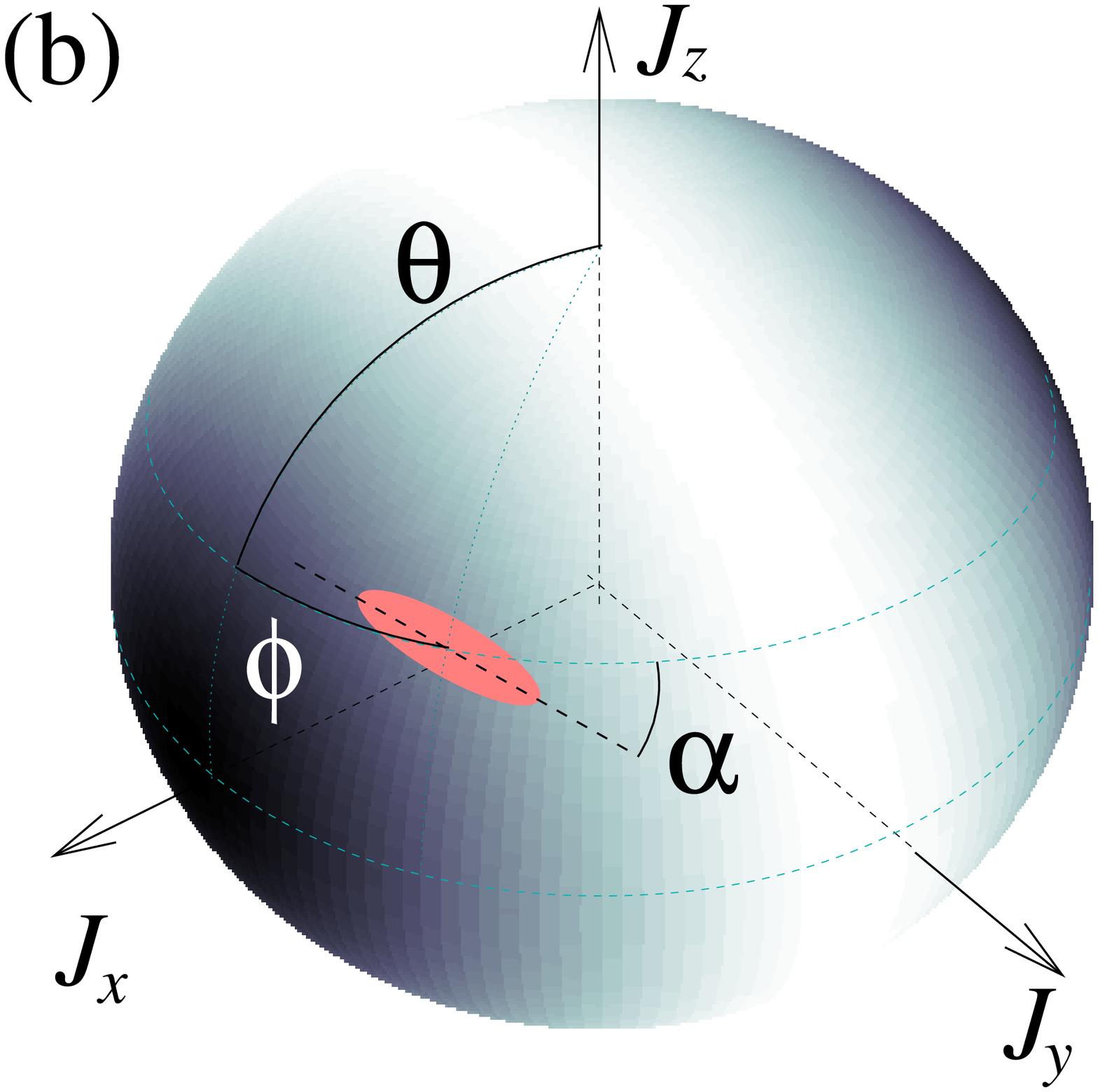,scale=0.18}}
\centerline{\epsfig{file=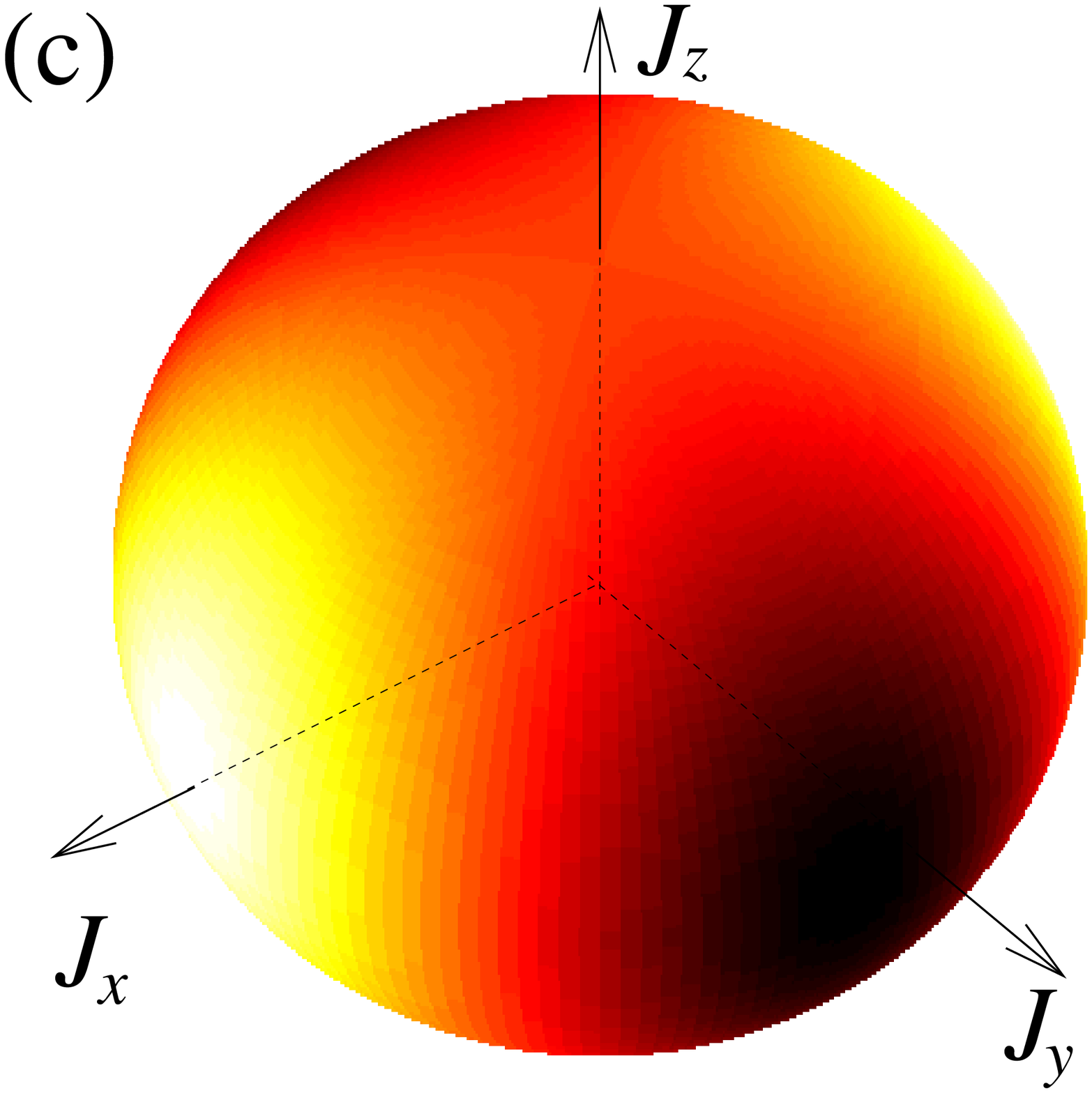,scale=0.18}  \quad \epsfig{file=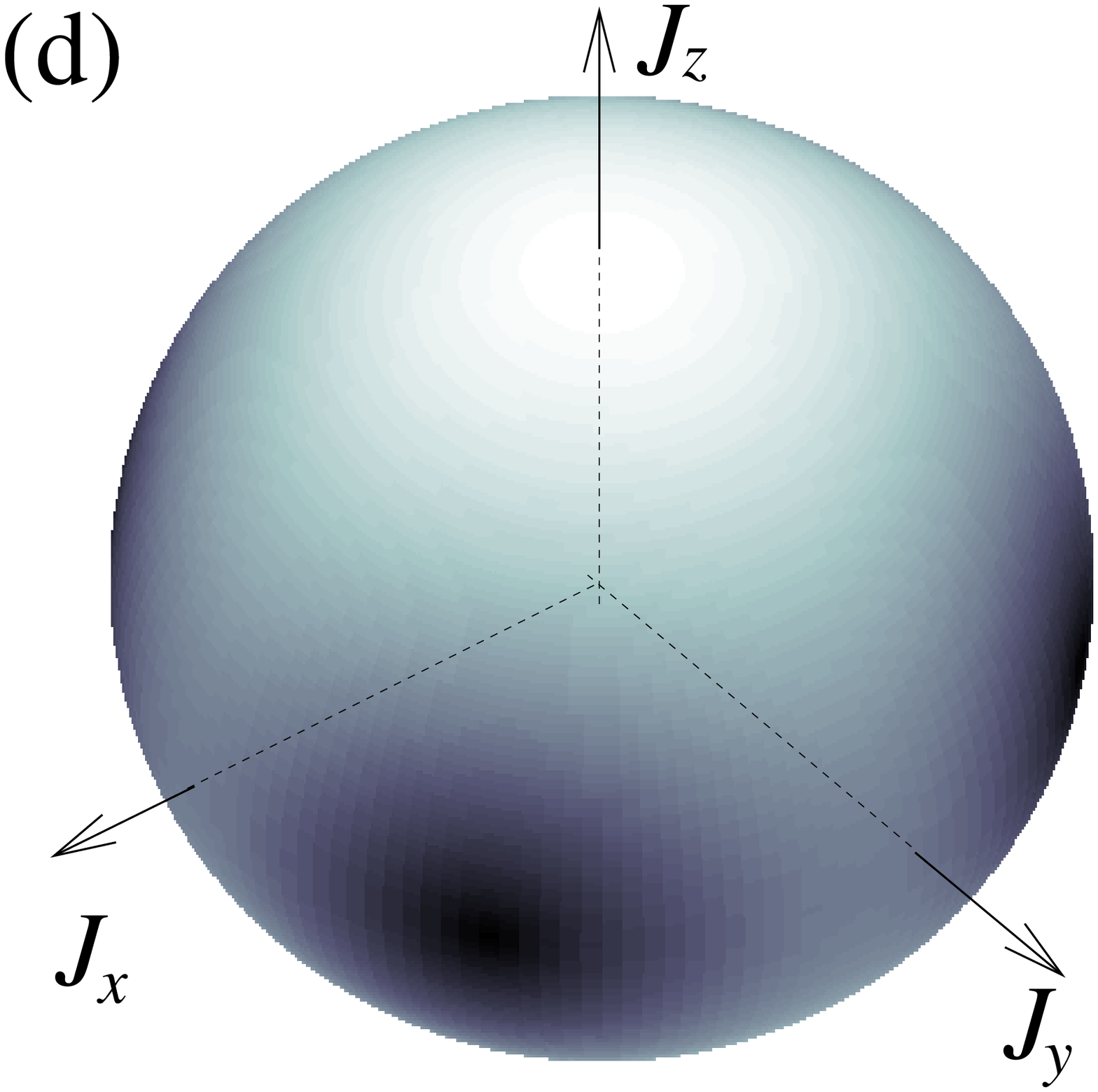,scale=0.18}}
\centerline{\epsfig{file=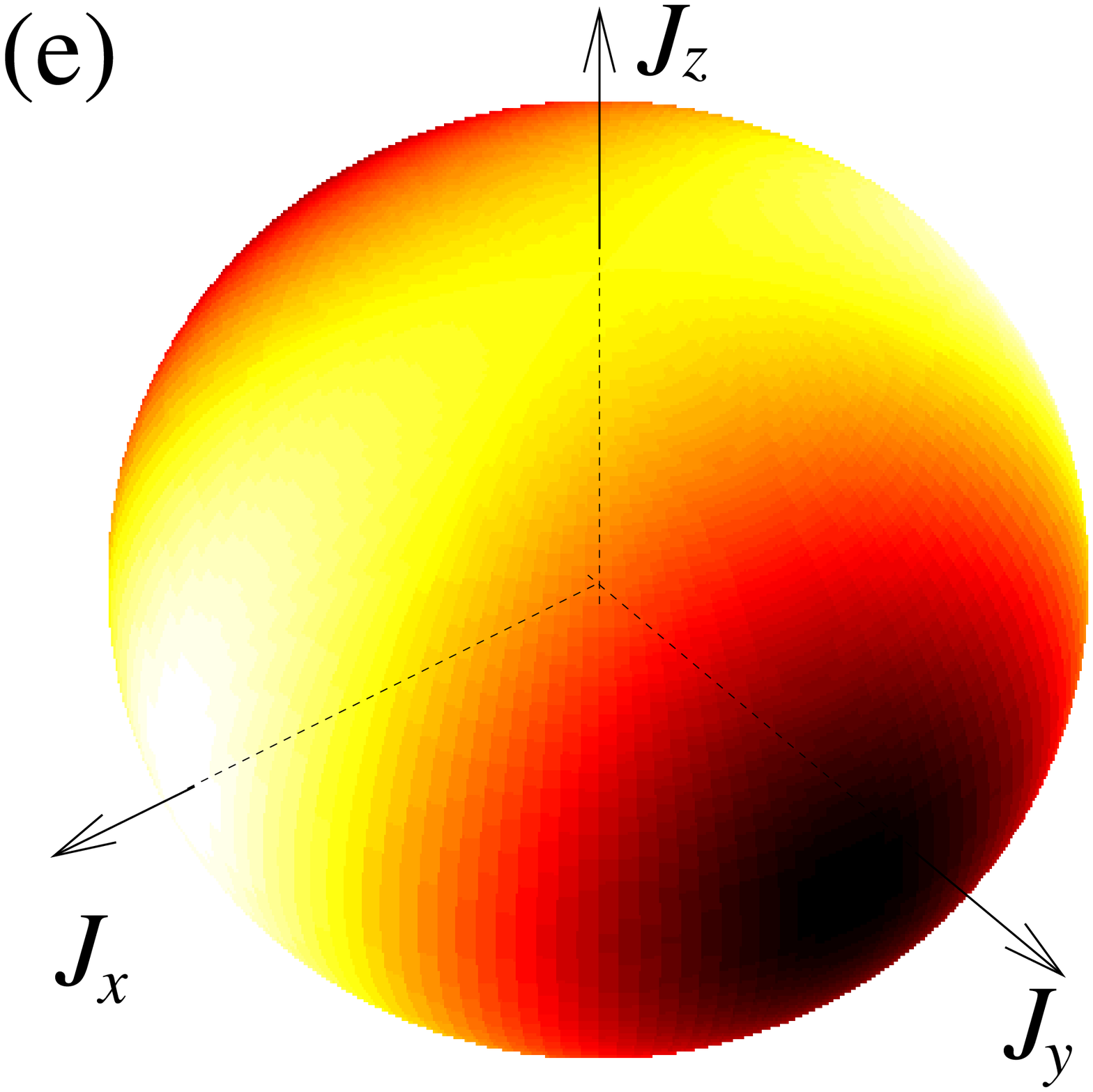,scale=0.18}  \quad \epsfig{file=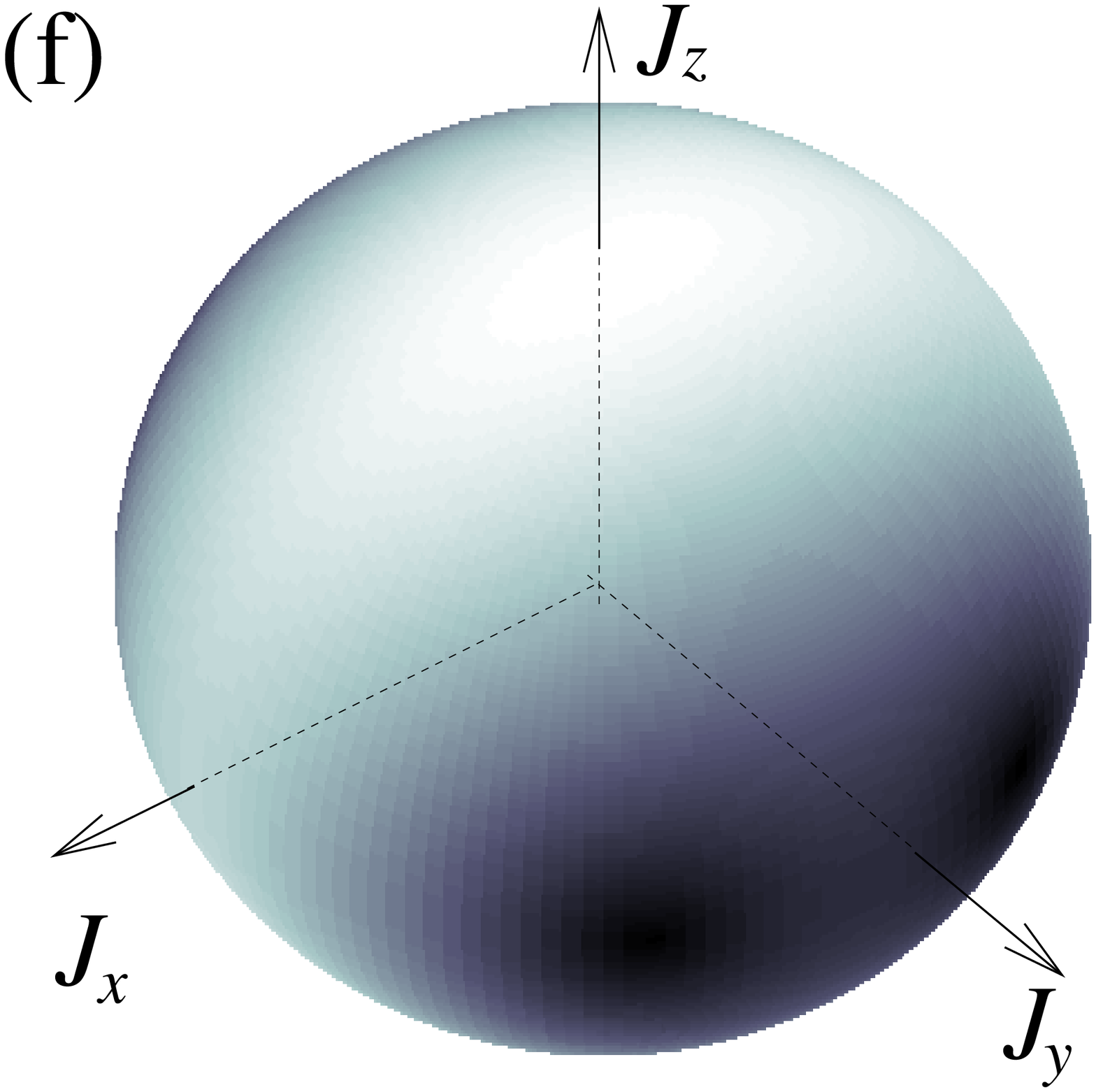,scale=0.18}}
\caption{\label{f-koule} (color online). Schematic pictures of the Hamiltonian $H$ (a, c, e) and squeezing rate $Q$ (b, d, f). The geometry of angles $\vartheta$, $\varphi$ and $\alpha$ is shown in (b). The shades correspond to the mean value of the Hamiltonian and to the squeezing rate of a spin coherent state with the direction of $\vec{J}$, the lighter shade corresponds to higher values. The Hamiltonian of Eq. (\ref{Ham}) is with $\vec{\omega} = 0$ and diagonal twisting tensor with $\{\chi_{xx},\chi_{yy},\chi_{zz}\} = \{1,0,0\}$ (OAT: a, b); $\{1,0,0.5\}$ (TACT: c, d); and $\{1,0,0.8\}$ (general twisting: e, f).}
\end{figure}

To find the maximum squeezing rate and optimum variance orientation for arbitrary location of the state, we have to transform the components of the twisting tensor.  For simplicity, we  choose the coordinate system oriented such that $\chi$ is diagonal. Two angles, $\vartheta$ and $\varphi$, determine the direction of the state as shown in Fig. \ref{f-koule}b. On calculating the elements of $\chi$ in the new coordinates one finds
\begin{eqnarray}
 Q &=& 2|{\cal J}| \left\{ \left[ \chi_x (\cos^2 \vartheta \cos^2 \varphi - \sin^2 \varphi)
\right. \right. \nonumber \\
& &+\left. \chi_y (\cos^2 \vartheta \sin^2 \varphi - \cos^2 \varphi) + \chi_z \sin^2 \vartheta \right]^2  \nonumber \\
& &+ \left. 4(\chi_x-\chi_y)^2\cos^2 \vartheta \cos^2 \varphi \sin^2 \varphi  \right\}^{1/2} ,
\label{rateQgen}
\end{eqnarray}
and 
\begin{eqnarray}
 \tan 2\alpha &=& \left[ \chi_x (\cos^2 \vartheta \cos^2 \varphi - \sin^2 \varphi)
\right.
\nonumber \\
& & \left. + \chi_y (\cos^2 \vartheta \sin^2 \varphi - \cos^2 \varphi) + \chi_z \sin^2 \vartheta \right] \nonumber \\
& & \times \left[ (\chi_y-\chi_x)\cos \vartheta \sin 2\varphi \right]^{-1},
\end{eqnarray}
where for the diagonal $\chi$ we used $\chi_{x}\equiv \chi_{xx}$, etc. 

From Eq. (\ref{rateQgen}) one can find for which directions $(\vartheta, \varphi)$ the squeezing rate is maximum and for which it is zero. Let us first consider the general case when the three eigenvalues of $\chi$ are all different (Fig. \ref{f-koule}c-f), e.g., $\chi_y < \chi_z < \chi_x$. Then there are four points where $Q=0$, all at the equator of the Bloch sphere, $\vartheta = \pi/2$, and $\sin^2 \varphi = (\chi_z-\chi_y)/(\chi_x-\chi_y)$. The maximum of $Q$ is achieved at the poles at $\vartheta=0$ and $\vartheta=\pi$ where $Q=2{\cal J}_z (\chi_x-\chi_y)$, and the optimum orientation of the squeezing ellipse is exactly half-way between the $J_x$ and $J_y$ directions. A special case of this situation is TACT with $\chi_x-\chi_z = \chi_z-\chi_y$ with symmetrical squeezing geometry (Fig. \ref{f-koule}c,d). Generally, the maximum achievable squeezing rate only depends on the difference between the maximum and minimum eigenvalues of the twisting tensor.

If $\chi$ is degenerate with, say $\chi_y=\chi_z$, $\chi_x > 0$ (OAT, Fig. \ref{f-koule}a,b), there are two zeros of $Q$ located at $\vartheta = \pi/2$, $\varphi =0, \pi$, and the maximum is achieved along the meridian  $\varphi = \pm \pi/2$ with $Q=2|{\cal J}| \chi_x$.

\section{Approximate solution of the equations of motion}
\label{secSolution}
For simplicity we choose the coordinate system such that the initial state is centered at the pole of the Bloch sphere (${\cal J}_x = {\cal J}_y =0$) oriented such that the off-diagonal term $\chi_{xy}$ vanishes. Let us further assume that the frequency components $\omega_x$ and $\omega_z$ are chosen such that the state is kept centered at the pole (this would be simply $\omega_{x,y} = 0$ if $\chi_{xz}=\chi_{yz}=0$, but otherwise a nontrivial expression for $\omega_{x,y}$ has to be used to compensate for the Bogoliubov backreaction). 
We introduce scaled variables $v_{kl}$, $j$ and $\tau$ as 
\begin{eqnarray}
V_{kl} &\equiv & \frac{N}{4}v_{kl}, \\
{\cal J}_z &\equiv & \frac{N}{2}j, \\
\omega_z &\equiv & N\tilde \omega, \\ 
t & \equiv & \frac{\tau}{N}, 
\end{eqnarray}
and get a closed set of equations
\begin{eqnarray}
\label{dvxxdt}
\frac{dv_{xx}}{d\tau} &=& 2\left[ -\tilde \omega + (\chi_{y}-\chi_{z}) j \right]v_{xy} , \\
\frac{dv_{yy}}{d\tau} &=& 2\left[ \tilde \omega - (\chi_{x}-\chi_{z}) j \right]v_{xy} , 
\label{dvyydt}
\\
\frac{dv_{xy}}{d\tau}  &=& \tilde \omega \left( v_{xx} -v_{yy} \right) 
\nonumber \\
& & + j \left[  (\chi_{z}-\chi_{x}) v_{xx} -   (\chi_{z}-\chi_{y}) v_{yy} \right] , 
\label{dvxydt}
\\
\frac{dj}{d\tau} &=& \frac{1}{N}(\chi_{x}-\chi_{y})v_{xy} .
\end{eqnarray}

\begin{figure}
\centerline{\epsfig{file=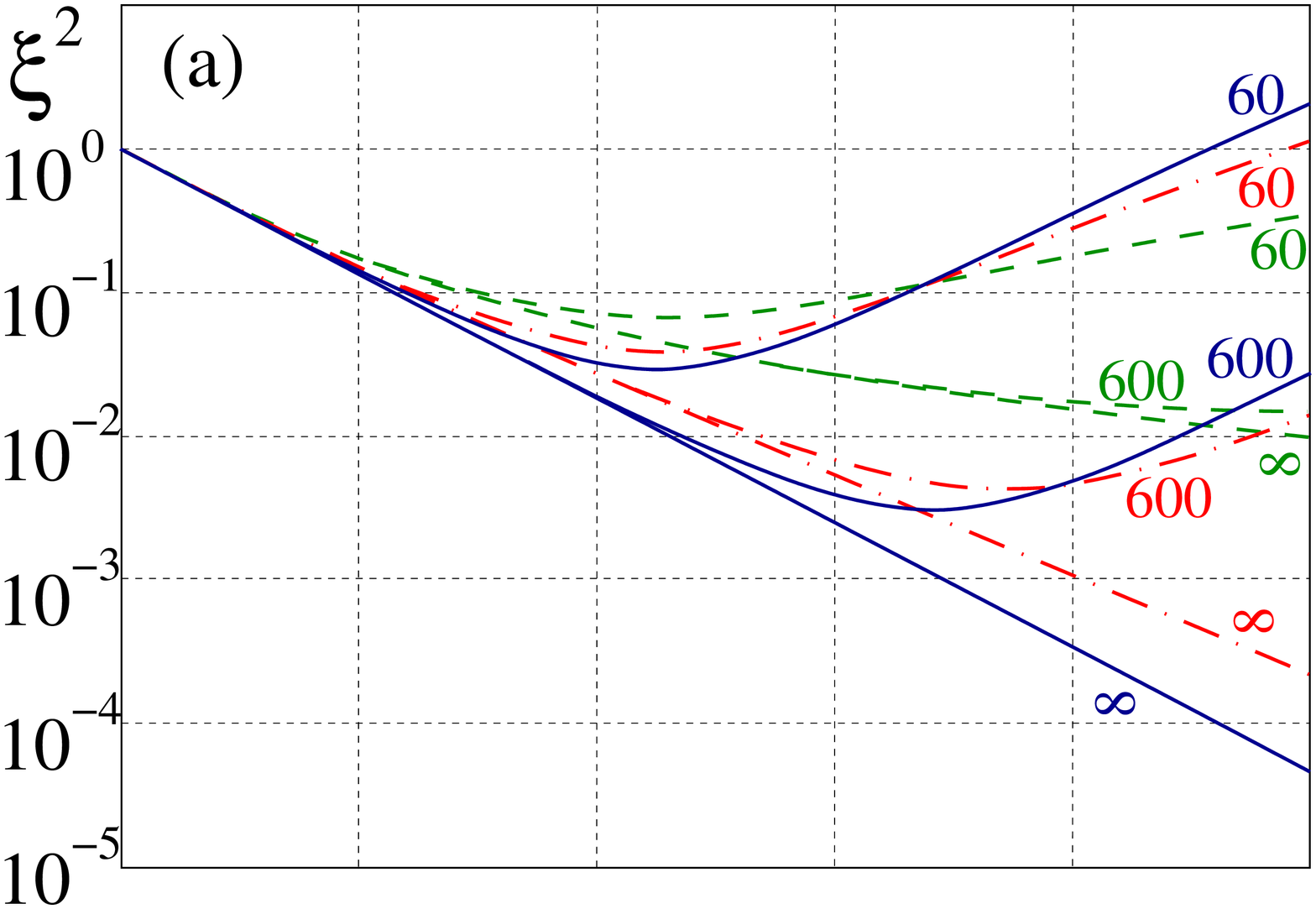,scale=0.26}}
\centerline{\epsfig{file=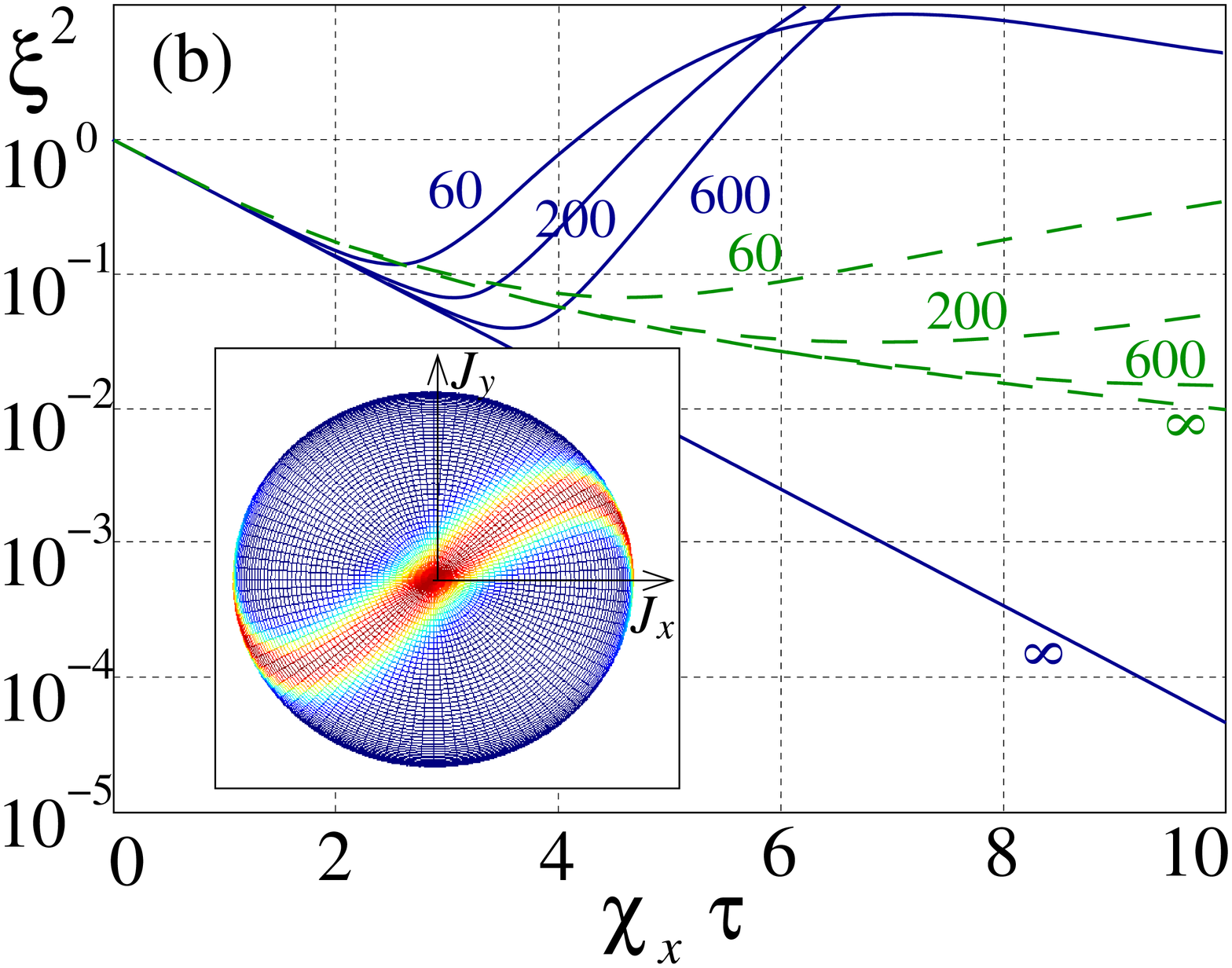,scale=0.26}}
\caption{\label{f-graf} (color online). Time evolution of the squeezing parameter, diagonal form of the twisting tensor with $\chi_x=1$ and $\chi_y=0$. (a) One axis squeezing with $\chi_z=0$ (green broken line), TACT with $\chi_z=0.5$ (full blue line), and general twisting with $\chi_z=0.8$ (red dash-dotted line), the number at each line being the number of particles $N$, the symbol $\infty$ denotes the asymptotics of Eq. (\ref{xi2gen}). (b) OAT with no rotation (green broken line), and with optimized rotation according to Eq. (\ref{tildeom}) (full blue line). Inset: Husimi function of the state with $N=60$, optimized rotation, and $\chi_x \tau=5$.}
\end{figure}

In the limit of $N\to \infty$, the derivative $dj/d\tau$ approaches zero and one can take $j=\pm 1$ and solve the equations with the initial condition of the spin coherent state $v_{xx}(0) = v_{yy}(0)=1$, $v_{xy}(0)=0$. 
 Let us assume $\chi_{x} \geq \chi_{z} \geq \chi_{y}$ and define 
\begin{eqnarray}
\Delta \chi_x &\equiv & \chi_{x}-\chi_{z} , \\ 
 \Delta \chi_y &\equiv& \chi_{z}-\chi_{y}, \\
\Delta \chi &\equiv& 2\sqrt{\Delta \chi_x \Delta \chi_y}. 
\end{eqnarray}
In the special case of $\tilde \omega=0$ we find on solving Eqs. (\ref{dvxxdt})--(\ref{dvxydt})
\begin{eqnarray}
\label{eqvxx}
 v_{xx} &=& \frac{1}{2\Delta \chi_x}\left[(\Delta \chi_x +\Delta \chi_y) \cosh (\Delta \chi \tau) \right. \nonumber \\
 & & \left. + \Delta \chi_x -\Delta \chi_y \right], \\
 v_{yy} &=& \frac{1}{2\Delta \chi_y}\left[(\Delta \chi_x +\Delta \chi_y) \cosh (\Delta \chi \tau) \right. \nonumber \\
 & & \left. + \Delta \chi_y -\Delta \chi_x \right], \\
v_{xy} &=& \frac{\Delta \chi_x +\Delta \chi_y}{2 \sqrt{\Delta \chi_x \Delta \chi_y}} \sinh (\Delta \chi \tau).
\label{eqvxy}
\end{eqnarray}
The squeezing parameter $\xi^2$ defined as the ratio of the minimum variance of the uncertainty ellipse of the state and the variance of the spin coherent state \cite{Wineland1994} is
\begin{eqnarray}
\xi^2 = \frac{1}{2}(v_{xx}+v_{yy})-\sqrt{v_{xy}^2+\frac{(v_{xx}-v_{yy})^2}{4}} .
\end{eqnarray}
Using the results of Eqs. (\ref{eqvxx})--(\ref{eqvxy}) we get
\begin{eqnarray}
 \xi^2 &=& \frac{1}{4\Delta \chi_x \Delta \chi_y} \left\{(\Delta \chi_x +\Delta \chi_y)^2 \cosh (\Delta \chi \tau) \right. \nonumber \\ 
& & - (\Delta \chi_x -\Delta \chi_y)^2
- \left[ (\Delta \chi_x +\Delta \chi_y)^4 \cosh ^2 (\Delta \chi \tau) \right. \nonumber \\ 
& & - 2(\Delta \chi_x -\Delta \chi_y)^2  (\Delta \chi_x +\Delta \chi_y)^2\cosh (\Delta \chi \tau) \nonumber \\ 
& & \left. \left. +(\Delta \chi_x -\Delta \chi_y)^4 - 16 \Delta \chi_x^2 \Delta \chi_y^2
\right] ^{1/2} \right\}.
\label{xi2gen}
\end{eqnarray}

The results are illustrated in Fig. \ref{f-graf}a. In the graphs, all lines were calculated by numerical solution of the Schr\"{o}dinger equation, the lines marked with symbol ``$\infty$'' coinciding with the analytical result of Eq. (\ref{xi2gen}).
Two special cases are worth mentioning, first, in the OAT with $\Delta \chi_y = 0$ (green broken line with symbol ``$\infty$'' in Fig. \ref{f-graf}a) the squeezing is
\begin{eqnarray}
\xi^2  = 1 - \Delta \chi_x \tau \sqrt{1 + \frac{\left( \Delta \chi_x \tau\right)^2}{4}}
+  \frac{1}{2}\left(\Delta \chi_x \tau\right)^2 .
\label{vpmoneax}
\end{eqnarray}
For short times this expression drops linearly as $1-\Delta \chi_x \tau$, and for long times it approaches zero as $1/(\Delta \chi_x \tau)^2$.
The second special case is TACT 
 with $\Delta \chi = 2\Delta \chi_x =  2\Delta \chi_y = \chi_{x}-\chi_{y}$ (blue full line with symbol ``$\infty$'' in Fig. \ref{f-graf}a) when we get
$ \xi^2  = \exp (-\Delta \chi \tau) $.
Although at the beginning for the two cases the squeezing evolves at the same rate given by the difference of the biggest and smallest eigenvalues of $\chi$, for longer times it drops to zero much faster in the  TACT case.  One should keep in mind that for longer times these results can only be used as long as the used approximations are valid (the deviation of the exact values for various finite $N$ from these approximate solutions can be seen in Fig  \ref{f-graf}a).

\section{Optimum rotation}
\label{secOptimum}
So far the special case of $\tilde \omega=0$ has been considered. 
To generate squeezing at the maximum rate, one has to keep the state optimally oriented with respect to the main twisting axes, so that $\alpha = \pi/4$  (see Eq. (\ref{tan2a}) with $\chi_{xy}=0$). This leads to 
\begin{eqnarray}
v_{xx}=v_{yy} = (v_{+}+v_{-})/2 
\end{eqnarray}
and 
\begin{eqnarray}
v_{xy} = (v_{+}-v_{-})/2.
\end{eqnarray} 
It follows that $dv_{xx}/d\tau = dv_{yy}/d\tau$, and from Eqs. (\ref{dvxxdt}) and (\ref{dvyydt}) that the optimum rotation frequency should satisfy
\begin{eqnarray}
\tilde \omega = j \left(\frac{\chi_{x}+\chi_{y}}{2}- \chi_{z} \right).
\label{tildeom}
\end{eqnarray}
Thus, for the exact TACT with $\chi_{z} = (\chi_{x}+\chi_{y})/2$ no rotation is needed to achieve the optimum squeezing rate. For any other values of the twisting parameters one needs to keep the variance ellipse optimally oriented by means of suitable rotation frequency.
The evolution then follows from Eqs. (\ref{dvxxdt})-(\ref{dvxydt}) as
\begin{eqnarray}
dv_{xx}/d\tau=dv_{yy}/d\tau = j(\chi_{y}-\chi_{x}) v_{xy}, 
\end{eqnarray}
and
\begin{eqnarray}
dv_{xy}/d\tau = j(\chi_{y}-\chi_{x}) v_{xx}.
\end{eqnarray}
If the system starts in the spin coherent state with $v_{xx}(0)=v_{yy}(0)=1$, and $v_{xy}(0)=0$, one finds 
\begin{eqnarray}
\xi^2 = \exp \left[- (\chi_{x}-\chi_{y})\tau \right]. 
\end{eqnarray}
These results are illustrated in Fig. 
\ref{f-graf}b. Note that the Gaussian approximation with finite $N$ works for relatively short times, after which the state undergoes an $S$-shape deformation and the squeezing is deteriorated (inset of  Fig. \ref{f-graf}b). These results hint for which parameters it might be suitable to apply the additional rotation during the squeezing preparation stage, e.g., the data of \cite{Gross2010} ($N\approx 400, \chi\tau \approx 3$) suggest a possible room for further optimization by this means.

\section{Conclusion} 
\label{secConcl}
The aim of this paper was to offer a unified tensor approach to all quadratic squeezing schemes that have so far been treated separately, such as one-axis twisting or two-axis counter-twisting. The main results are derivation of a closed set of equations governing the first and second moments in Gaussian approximation, and 
showing the role of eigenvalues of the twisting tensor. At the early stages of squeezing, the most relevant parameter is the difference between the maximum and minimum eigenvalues of the twisting tensor which determines the squeezing rate. For longer times, most efficient squeezing is achieved if the middle eigenvalue halves the interval between the extreme ones (TACT). In other cases, for certain times the imbalance of the middle eigenvalue can be compensated by suitable rotation.

The approach is suitable for two-mode systems with quadratic nonlinearities, such as two-mode optical resonators with Kerr media, BEC in structured traps, collective atomic spins, etc. Apart from covering various squeezing scenarios important mostly for quantum metrology and interferometry, it is also relevant for the LMG model studied as a paradigm of quantum phase transitions. Further generalization to cover losses and decoherence \cite{Losses} and various squeezing optimization strategies will be the subject of a forthcoming work. 

\acknowledgments

Stimulating discussions with K. K. Das, M. Kol\'{a}\v{r}, and K. M\o{}lmer are acknowledged. This work was supported by the European Social Fund and the state budget of the Czech Repuplic, project  CZ.1.07/2.3.00/30.0041.

\end{document}